\documentclass[sigconf, authorversion, nonacm]{acmart}
\usepackage{enumitem}
\usepackage{caption}
\usepackage{subcaption}
\usepackage{float}
\usepackage{multirow}
\usepackage{amsmath}
\usepackage{cases}
\usepackage{array}
\usepackage{tikz}
\usepackage{quantikz}
\usepackage{amsmath}

\newcommand{\rowSep}{0.25cm}
\newcommand{\colSep}{0.2cm}

\newcommand{\myCnot}{
\begin{quantikz} [row sep= \rowSep, column sep= \colSep]
	\lstick{$a$} & \ctrl{1} & \qw & \rstick{$a$}\\
	\lstick{$b$} & \targ{}  & \qw & \rstick{$a \oplus b$}
\end{quantikz}
}

\newcommand{\myToffoli}{
\begin{quantikz} [row sep= \rowSep, column sep= \colSep]
	\lstick{$a$} & \ctrl{1} & \qw & \rstick{$a$} \\
	\lstick{$b$} & \ctrl{1} & \qw & \rstick{$b$} \\
	\lstick{$c$} & \targ{}  & \qw & \rstick{$a \cdot b \oplus c$}
\end{quantikz}
}

\newcommand{\KimModuloAdderBoxed}{
\begin{quantikz} [row sep= 0.2cm, column sep= 0.17cm, transparent]
	\lstick{$0$}   & \targ{} &\qw	& \qw\gategroup[13,steps=9,style={blue!75!black, ultra thick, dashed, rounded corners, inner xsep=2pt}, background]{{\sc QRCA based Carry Generation Logic}}
												& \qw		& \ctrl{2}	& \qw		& \qw		& \qw		& \qw		& \qw		& \ctrl{2}	& \qw  & \targ{}	& \targ{}	& \qw	& \ctrl{2}\gategroup[13,steps=19,style={green!50!black, ultra thick, dashed, rounded corners, inner xsep=2pt}, background]{{\sc QRCA based Carryout Truncated Full Adder}}		
																																																	& \qw		& \qw		& \qw		& \qw		& \qw		& \qw		& \qw		& \qw		& \qw		& \qw		& \qw		& \qw		& \qw		& \ctrl{2}	& \qw		& \qw		& \qw		& \ctrl{2}		& \qw  & \targ{}	& \qw  & \rstick{$0$} \\
	\lstick{$a_0$} & \qw  	 &\qw	& \ctrl{1} 	& \ctrl{1} 	& \qw 		& \qw 		& \qw 		& \qw 		& \qw 		& \qw 		&  \qw 	 	& \qw  & \qw 		& \qw 		& \qw	& \qw 		& \qw 		& \qw 		& \qw 		& \qw 		& \qw 		& \qw 	 	& \qw 		& \qw 		& \qw 		& \qw 		& \qw 		& \qw 		& \qw 		&  \qw 	  	& \ctrl{1} 	& \ctrl{2} 	& \ctrl{1} 	& \qw 		 	& \qw  & \qw 		& \qw  & \rstick{$a_0$} \\
	\lstick{$b_0$} & \qw  	 &\qw	& \ctrl{1}	& \targ{}	& \ctrl{1}	& \qw		& \qw		& \qw		& \qw		& \qw		& \ctrl{1}	& \qw  & \qw		& \qw		& \qw	& \ctrl{1}	& \qw		& \qw		& \qw		& \qw		& \qw		& \qw		& \qw		& \qw		& \qw		& \qw		& \qw		& \qw		& \qw		& \ctrl{1} 	& \targ{} 	& \ctrl{1} 	& \targ{} 	& \targ{} 		& \qw  & \qw		& \qw  & \rstick{$s_0$} \\
	\lstick{$0$}   & \qw  	 &\qw	& \targ{}	& \qw		& \targ{}	& \ctrl{2}	& \qw		& \qw		& \qw		& \ctrl{2}	& \targ{}	& \qw  & \qw		& \qw		& \qw	& \targ{}	& \ctrl{2}	& \qw		& \qw		& \qw		& \qw		& \qw		& \qw		& \qw		& \ctrl{2}	& \qw		& \qw		& \qw		& \ctrl{2}	& \targ{}	& \qw		& \targ{}	& \qw		& \qw			& \qw  & \qw		& \qw  & \rstick{$c_1$} \\
	\lstick{$a_1$} & \qw  	 &\qw	& \ctrl{1} 	& \ctrl{1} 	& \qw 		&  \qw 	 	& \qw 		& \qw 		& \qw 		&  \qw 	 	& \qw 		& \qw  & \qw 		& \qw 		& \qw	& \qw 		&  \qw 	 	& \qw 		& \qw 		& \qw 		& \qw 		& \qw 	 	& \qw 		& \qw 		&  \qw 	  	& \ctrl{1} 	& \ctrl{2} 	& \ctrl{1} 	& \qw 		& \qw 		& \qw	 	& \qw		& \qw		& \qw 			& \qw  & \qw 		& \qw  & \rstick{$a_1$} \\
	\lstick{$b_1$} & \qw  	 &\qw	& \ctrl{1} 	& \targ{} 	& \qw 		& \ctrl{1} 	& \qw 		& \qw 		& \qw 		& \ctrl{1} 	& \qw 		& \qw  & \qw 		& \qw 		& \qw	& \qw 		& \ctrl{1} 	& \qw 		& \qw 		& \qw 		& \qw 		& \qw 	 	& \qw 		& \qw 		& \ctrl{1} 	& \targ{} 	& \ctrl{1} 	& \targ{} 	& \targ{} 	& \qw 		& \qw	 	& \qw		& \qw		& \qw 			& \qw  & \qw 		& \qw  & \rstick{$s_1$} \\
	\lstick{$0$}   & \qw  	 &\qw	& \targ{}	& \qw		& \qw		& \targ{}	& \ctrl{2}	& \qw		& \ctrl{2}	& \targ{}	& \qw		& \qw  & \qw		& \qw		& \qw	& \qw		& \targ{}	& \ctrl{2}	& \qw		& \ctrl{2}	& \qw		& \qw		& \qw		& \ctrl{2}	& \targ{}	& \qw		& \targ{}	& \qw		& \qw		& \qw		& \qw		& \qw		& \qw		& \qw			& \qw  & \qw		& \qw  & \rstick{$c_2$} \\
	\lstick{$a_2$} & \qw  	 &\qw	& \ctrl{1} 	& \ctrl{1} 	& \qw 		& \qw 		&  \qw 	 	& \qw 		&  \qw 	 	& \qw 		& \qw 		& \qw  & \qw 		& \qw 		& \qw	& \qw 		& \qw 		&  \qw 	 	& \qw 		&  \qw 	  	& \ctrl{1} 	& \ctrl{2} 	& \ctrl{1} 	& \qw 		& \qw 		& \qw	 	& \qw 		& \qw	 	& \qw 		& \qw 		& \qw	 	& \qw 		& \qw	 	& \qw 			& \qw  & \qw 		& \qw  & \rstick{$a_2$} \\
	\lstick{$b_2$} & \qw  	 &\qw	& \ctrl{1} 	& \targ{} 	& \qw 		& \qw 		& \ctrl{1} 	& \qw 		& \ctrl{1} 	& \qw 		& \qw 		& \qw  & \qw 		& \qw 		& \qw	& \qw 		& \qw 		& \ctrl{1} 	& \qw 		& \ctrl{1} 	& \targ{} 	& \ctrl{1} 	& \targ{} 	& \targ{} 	& \qw 		& \qw	 	& \qw 		& \qw	 	& \qw 		& \qw 		& \qw	 	& \qw 		& \qw	 	& \qw 			& \qw  & \qw 		& \qw  & \rstick{$s_2$} \\
	\lstick{$0$}   & \qw  	 &\qw	& \targ{}	& \qw		& \qw		& \qw		& \targ{}	& \ctrl{2}	& \targ{}	& \qw		& \qw		& \qw  & \qw		& \qw		& \qw	& \qw		& \qw		& \targ{}	& \ctrl{2}	& \targ{}	& \qw		& \targ{}	& \qw		& \qw		& \qw		& \qw		& \qw		& \qw		& \qw		& \qw		& \qw		& \qw		& \qw		& \qw			& \qw  & \qw		& \qw  & \rstick{$c_3$} \\
	\lstick{$a_3$} & \qw  	 &\qw	& \ctrl{1} 	& \ctrl{1} 	& \qw 		& \qw 		& \qw 		&  \qw 	 	& \qw 		& \qw 		& \qw 		& \qw  & \qw 		& \qw 		& \qw	& \qw 		& \qw 		& \qw 		& \qw 	 	& \qw 		& \qw 		& \qw 		& \qw 		& \qw 		& \qw 		& \qw	 	& \qw 		& \qw	 	& \qw 		& \qw 		& \qw	 	& \qw 		& \qw	 	& \qw 			& \qw  & \qw 		& \qw  & \rstick{$a_3$} \\
	\lstick{$b_3$} & \qw  	 &\qw	& \ctrl{1} 	& \targ{} 	& \qw 		& \qw 		& \qw 		& \ctrl{1} 	& \qw 		& \qw 		& \qw 		& \qw  & \qw 		& \qw 		& \qw	& \qw 		& \qw 		& \qw 		& \targ{} 	& \qw 		& \qw 		& \qw 		& \qw 		& \qw 		& \qw 		& \qw	 	& \qw 		& \qw	 	& \qw 		& \qw 		& \qw	 	& \qw 		& \qw	 	& \qw 			& \qw  & \qw 		& \qw  & \rstick{$s_3$} \\
	\lstick{$0$}   & \qw  	 &\qw	& \targ{} 	& \qw 		& \qw 		& \qw 		& \qw 		& \targ{} 	& \qw 		& \qw 		& \qw 		& \qw  & \ctrl{-12}& \qw 		& \qw	& \qw 		& \qw 		& \qw 		& \qw 		& \qw 		& \qw 		& \qw 		& \qw 		& \qw 		& \qw 		& \qw	 	& \qw 		& \qw	 	& \qw 		& \qw 		& \qw	 	& \qw 		& \qw	 	& \qw 			& \qw  & \ctrl{-12}& \qw  & \rstick{$c_4$}
\end{quantikz}
}

\newcommand{\ProposedModuloAdderBoxed}{
\begin{quantikz} [row sep= \rowSep, column sep= \colSep, transparent]
	\lstick{$1$}   & \qw		& \qw\gategroup[16,steps=7,style={blue!75!black, ultra thick, dashed, rounded corners, inner xsep=2pt}, background]{{\sc Carry-lookahead Generation Logic}}		
											& \qw		& \ctrl{3}	& \qw		& \qw		& \qw		& \qw		& \qw		& \targ{}	& \targ{}	& \qw		& \qw\gategroup[16,steps=14,style={green!50!black, ultra thick, dashed, rounded corners, inner xsep=2pt}, background]{{\sc Carry-lookahead Truncated Full Adder}}
																																												& \qw		& \ctrl{4}	& \qw		& \qw		& \ctrl{2}	& \qw		& \qw		& \qw		& \qw		& \ctrl{4}	& \qw		& \qw		& \qw		& \qw		& \targ{}	& \qw		& \rstick{$0$} \\
	\lstick{$a_0$} & \qw		& \ctrl{2}	& \qw		& \qw		& \qw		& \qw		& \qw		& \ctrl{1}	& \qw		& \qw		& \qw		& \qw		& \ctrl{3}	& \ctrl{1}	& \qw		& \qw		& \qw		& \qw		& \qw		& \ctrl{1}	& \qw		& \qw		& \qw		& \ctrl{1}	& \ctrl{3}	& \qw		& \qw		& \qw		& \qw		& \rstick{$a_0$} \\
	\lstick{$b_0$} & \qw		& \ctrl{1}	& \ctrl{1}	& \ctrl{1}	& \qw		& \qw		& \qw		& \targ{}	& \qw		& \qw		& \qw		& \qw		& \ctrl{2}	& \targ{}	& \ctrl{2}	& \qw		& \qw		& \targ{}	& \targ{}	& \targ{}	& \qw		& \qw		& \ctrl{2}	& \targ{}	& \ctrl{2}	& \targ{}	& \qw		& \qw		& \qw		& \rstick{$s_0$} \\
	\lstick{$0$}   & \qw		& \targ{}	& \targ{}	& \targ{}	& \ctrl{4}	& \qw		& \qw		& \qw		& \qw		& \qw		& \qw		& \qw		& \qw		& \qw		& \qw		& \qw		& \qw		& \qw		& \qw		& \qw		& \qw		& \qw		& \qw		& \qw		& \qw		& \qw		& \qw		& \qw		& \qw		& \rstick{$c_1$} \\
	\lstick{$0$}   & \qw		& \qw		& \qw		& \qw		& \qw		& \qw		& \qw		& \qw		& \qw		& \qw		& \qw		& \qw		& \targ{}	& \qw		& \targ{}	& \ctrl{4}	& \qw		& \ctrl{2}	& \qw		& \qw		& \qw		& \ctrl{4}	& \targ{}	& \qw		& \targ{}	& \qw		& \qw		& \qw		& \qw		& \rstick{$0$} \\
	\lstick{$a_1$} & \qw		& \ctrl{2}	& \ctrl{1}	& \qw		& \qw		& \qw		& \qw		& \ctrl{1}	& \qw		& \qw		& \qw		& \qw		& \ctrl{3}	& \ctrl{1}	& \qw		& \qw		& \qw		& \qw		& \qw		& \ctrl{1}	& \qw		& \qw		& \qw		& \ctrl{1}	& \ctrl{3}	& \qw		& \qw		& \qw		& \qw		& \rstick{$a_1$} \\
	\lstick{$b_1$} & \qw		& \ctrl{1}	& \targ{}	& \qw		& \ctrl{1}	& \qw		& \qw		& \targ{}	& \qw		& \qw		& \qw		& \qw		& \ctrl{2}	& \targ{}	& \qw		& \ctrl{2}	& \qw		& \targ{}	& \targ{}	& \targ{}	& \qw		& \ctrl{2}	& \qw		& \targ{}	& \ctrl{2}	& \targ{}	& \qw		& \qw		& \qw		& \rstick{$s_1$} \\
	\lstick{$0$}   & \qw		& \targ{}	& \qw		& \qw		& \targ{}	& \ctrl{8}	& \qw		& \qw		& \qw		& \qw		& \qw		& \qw		& \qw		& \qw		& \qw		& \qw		& \qw		& \qw		& \qw		& \qw		& \qw		& \qw		& \qw		& \qw		& \qw		& \qw		& \qw		& \qw		& \qw		& \rstick{$c_2$} \\
	\lstick{$0$}   & \qw		& \qw		& \qw		& \qw		& \qw		& \qw		& \qw		& \qw		& \qw		& \qw		& \qw		& \qw		& \targ{}	& \qw		& \qw		& \targ{}	& \ctrl{3}	& \ctrl{2}	& \qw		& \qw		& \ctrl{3}	& \targ{}	& \qw		& \qw		& \targ{}	& \qw		& \qw		& \qw		& \qw		& \rstick{$0$} \\
	\lstick{$a_2$} & \qw		& \ctrl{3}	& \ctrl{1}	& \qw		& \qw		& \qw		& \qw		& \ctrl{1}	& \qw		& \qw		& \qw		& \qw		& \ctrl{2}	& \ctrl{1}	& \qw		& \qw		& \qw		& \qw		& \qw		& \ctrl{1}	& \qw		& \qw		& \qw		& \ctrl{1}	& \ctrl{2}	& \qw		& \qw		& \qw		& \qw		& \rstick{$a_2$} \\
	\lstick{$b_2$} & \qw		& \ctrl{2}	& \targ{}	& \ctrl{1}	& \qw		& \qw		& \ctrl{1}	& \targ{}	& \qw		& \qw		& \qw		& \qw		& \ctrl{1}	& \targ{}	& \qw		& \qw		& \ctrl{1}	& \targ{}	& \targ{}	& \targ{}	& \ctrl{1}	& \qw		& \qw		& \targ{}	& \ctrl{1}	& \targ{}	& \qw		& \qw		& \qw		& \rstick{$s_2$} \\
	\lstick{$0$}   & \qw		& \qw		& \qw		& \targ{}	& \qw		& \ctrl{4}	& \targ{}	& \qw		& \qw		& \qw		& \qw		& \qw		& \targ{}	& \qw		& \qw		& \qw		& \targ{}	& \ctrl{3}	& \qw		& \qw		& \targ{}	& \qw		& \qw		& \qw		& \targ{}	& \qw		& \qw		& \qw		& \qw		& \rstick{$0$} \\
	\lstick{$0$}   & \qw		& \targ{}	& \qw		& \qw		& \ctrl{3}	& \qw		& \qw		& \qw		& \qw		& \qw		& \qw		& \qw		& \qw		& \qw		& \qw		& \qw		& \qw		& \qw		& \qw		& \qw		& \qw		& \qw		& \qw		& \qw		& \qw		& \qw		& \qw		& \qw		& \qw		& \rstick{$c_3$} \\
	\lstick{$a_3$} & \qw		& \ctrl{2}	& \ctrl{1}	& \qw		& \qw		& \qw		& \qw		& \ctrl{1}	& \qw		& \qw		& \qw		& \qw		& \qw		& \ctrl{1}	& \qw		& \qw		& \qw		& \qw		& \qw		& \ctrl{1}	& \qw		& \qw		& \qw		& \ctrl{1}	& \qw		& \qw		& \qw		& \qw		& \qw		& \rstick{$a_3$} \\
	\lstick{$b_3$} & \qw		& \ctrl{1}	& \targ{}	& \ctrl{-3}	& \ctrl{1}	& \qw		& \ctrl{-3}	& \targ{}	& \qw		& \qw		& \qw		& \qw		& \qw		& \targ{}	& \qw		& \qw		& \qw		& \targ{}	& \targ{}	& \targ{}	& \qw		& \qw		& \qw		& \targ{}	& \qw		& \targ{}	& \qw		& \qw		& \qw		& \rstick{$s_3$} \\
	\lstick{$0$}   & \qw		& \targ{}	& \qw		& \qw		& \targ{}	& \targ{}	& \qw		& \qw		& \qw		& \ctrl{-15}& \qw		& \qw		& \qw		& \qw		& \qw		& \qw		& \qw		& \qw		& \qw		& \qw		& \qw		& \qw		& \qw		& \qw		& \qw		& \qw		& \qw		& \ctrl{-15}& \qw		& \rstick{$c_4$}
\end{quantikz}
}

\AtBeginDocument{%
  \providecommand\BibTeX{{%
    \normalfont B\kern-0.5em{\scshape i\kern-0.25em b}\kern-0.8em\TeX}}}

\newcommand{\orcidnew}[1]{\href{https://orcid.org/#1}{\includegraphics[width=8pt]{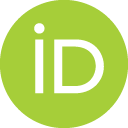}}}
\begin{document}

%%
%% The "title" command has an optional parameter,
%% allowing the author to define a "short title" to be used in page headers.
\title{A Logarithmic Depth\\Quantum Carry-Lookahead Modulo (2\textsuperscript{n} - 1) Adder}

%%
%% The "author" command and its associated commands are used to define
%% the authors and their affiliations.
%% Of note is the shared affiliation of the first two authors, and the
%% "authornote" and "authornotemark" commands
%% used to denote shared contribution to the research.
%\author{Bhaskar Gaur$^{1}$ \orcidnew{0000-0001-6738-6890}, Edgard Mu\~{n}oz-Coreas$^{2}$ and Himanshu Thapliyal$^{1}$ \orcidnew{0000-0001-9157-4517}\\ \textit{$^{1}$University of Tennessee, Knoxville, TN, USA;} \textit{$^{2}$University of North Texas, Denton, TX, USA} }

\author{Bhaskar Gaur \orcidnew{0000-0001-6738-6890}}
\affiliation{%
	\institution{University of Tennessee}
	\city{Knoxville}
	\state{TN}
	\country{USA}}
\email{bgaur@vols.utk.edu}
%\orcid{0000-0001-6738-6890}

\author{ Edgard Mu\~{n}oz-Coreas \orcidnew{0009-0008-9822-6724}}
\affiliation{%
	\institution{University of North Texas}
	\city{Denton}
	\state{TX}
	\country{USA}}
\email{Edgard.Munoz-coreas@unt.edu}

\author{Himanshu Thapliyal \orcidnew{0000-0001-9157-4517}}
\affiliation{%
	\institution{University of Tennessee}
	\city{Knoxville}
	\state{TN}
	\country{USA}}
\email{hthapliyal@utk.edu}

%%
%% By default, the full list of authors will be used in the page
%% headers. Often, this list is too long, and will overlap
%% other information printed in the page headers. This command allows
%% the author to define a more concise list
%% of authors' names for this purpose.
\renewcommand{\shortauthors}{Bhaskar Gaur, Edgard Mu\~{n}oz-Coreas, \& Himanshu Thapliyal} %% No italics

%%
%% The abstract is a short summary of the work to be presented in the
%% article.
\vspace{3mm}
\begin{abstract}
  Quantum Computing is making significant advancements toward creating machines capable of implementing quantum algorithms in various fields, such as quantum cryptography, quantum image processing, and optimization. The development of quantum arithmetic circuits for modulo addition is vital for implementing these quantum algorithms. While it is ideal to use quantum circuits based on fault-tolerant gates to overcome noise and decoherence errors, the current Noisy Intermediate Scale Quantum (NISQ) era quantum computers cannot handle the additional computational cost associated with fault-tolerant designs. Our research aims to minimize circuit depth, which can reduce noise and facilitate the implementation of quantum modulo addition circuits on NISQ machines. This work presents quantum carry-lookahead modulo (2\textsuperscript{n} - 1) adder (QCLMA), which is designed to receive two n-bit numbers and perform their addition with an O(log n) depth. Compared to existing work of O(n) depth, our proposed QCLMA reduces the depth and helps increase the noise fidelity. In order to increase error resilience, we also focus on creating a tree structure based Carry path, unlike the chain based Carry path of the current work. We run experiments on Quantum Computer IBM Cairo to evaluate the performance of the proposed QCLMA against the existing work and define Quantum State Fidelity Ratio (QSFR) to quantify the closeness of the correct output to the top output. When compared against existing work, the proposed QCLMA achieves a 47.21\% increase in QSFR for 4-qubit modulo addition showcasing its superior noise fidelity.

\end{abstract}

%%
%% The code below is generated by the tool at http://dl.acm.org/ccs.cfm.
%% Please copy and paste the code instead of the example below.
%%
%%
%% Keywords. The author(s) should pick words that accurately describe
%% the work being presented. Separate the keywords with commas.
\keywords{quantum adders, modulo addition, noise, carry-lookahead, quantum circuit, quantum computing, NISQ, FTQ}

%%
%% This command processes the author and affiliation and title
%% information and builds the first part of the formatted document.
\maketitle

\section*{Acknowledgement}
This research used resources of the Oak Ridge Leadership Computing Facility, which is a DOE Office of Science User Facility supported under Contract DE-AC05-00OR22725.

\section{Introduction}
\label{Introduction}
Quantum computing is an upcoming field with many potential applications, including cryptography, optimization, machine learning, communication, and simulating scientific experiments. Quantum arithmetic circuits enable efficient implementations of quantum computing algorithms such as variational quantum algorithms, quantum approximate optimization algorithms, singular value thresholding, HHL algorithm, elliptic curve cryptography (ECC), Shor's algorithm, and quantum convolution neural networks (QCNN) \cite{shor1994algorithms, ding2005rainbow, hur2022quantum, lee2019hybrid}.

Quantum modulo arithmetic circuits are a subset of the quantum arithmetic circuits that perform arithmetic operations by returning the remainder of the division of the result by a given modulus. The result of a modulo operation is always within a finite range (0, modulus - 1), which ensures closure under addition, subtraction, and multiplication. Among various quantum modulo operations, quantum modulo adder is the most basic and versatile quantum circuit, which can help create quantum modulo arithmetic circuits for subtraction, multiplication and exponentiation \cite{thapliyal2016mapping, rodney2005, wu2020novel}. In addition, quantum modulo adders fulfill specialized purposes in quantum image processing (QIP), quantum cryptography, and quantum convolution neural networks by reducing the cost of computations and operating in a finite set of numbers \cite{shor1994algorithms, ding2005rainbow, petzoldt2020efficient, jiang2014quantum, hur2022quantum, wu2020novel}.
%(quantum modular multiplication circuit cho et al)

Vedral et al. designed one of the earliest known general modulo N adders \cite{vedral1996quantum}. This adder generates a modulo sum by adding the two inputs and subtracting N using a simple mechanism. The final output depends on whether the Sum is larger or smaller than N. However, this design includes five stages that increase the quantum gate count and depth, making it impractical for the NISQ era.

Therefore, upcoming research endeavors focus on designing resource-efficient modulo adders that can be applied to crucial moduli with practical significance in designated fields. For example, the quantum modulo 2\textsuperscript{n} adder is utilized in quantum image scrambling and quantum image encryption \cite{jiang2014quantum}. Most quantum modulo 2\textsuperscript{n} adders are effortlessly derived from existing quantum full adder designs by carryout truncation \cite{cuccaro2004new, takahashi2005linear}. The quantum modulo (2\textsuperscript{n} - 1) adder, on the other hand, reduces computational cost in convolution neural networks and assumes a crucial role in public cryptographic systems that operate on moduli expressible in Galois Field GF(2\textsuperscript{n} - 1) \cite{hur2022quantum, ding2005rainbow, petzoldt2020efficient}. Nevertheless, unlike quantum modulo 2\textsuperscript{n} adder designs, designing a quantum modulo (2\textsuperscript{n} - 1) adder is not a straightforward undertaking, with the design proposed by Kim et al. currently representing the most efficient option \cite{kim2021quantum}.

Unlike Vedral et al., the design proposed by Kim et al. employs only two stages. Nevertheless, it still employs quantum ripple-carry addition (QRCA) for generating those two stages and consequently inherits the shortcomings of QRCA \cite{vedral1996quantum, kim2021quantum, cuccaro2004new}. The QRCA logic entails a lengthy Carry path, which increases the carryout's depth and noise susceptibility. Additionally, the output's least significant bits (LSB) are generated last, despite being inputted first, leading to increased idle time and a higher probability of being impacted by amplitude damping and depolarizing noise sources \cite{jayashankar2022achieving}.

In order to address these issues, we propose the use of a quantum carry-lookahead modulo (2\textsuperscript{n} - 1) adder (QCLMA), which has an O(log n) depth in contrast to Kim et al.'s quantum modulo (2\textsuperscript{n} - 1) adder, which has an O(n) depth. QCLMA's lower depth, when compared to Kim et al.'s version, leads to a reduction in idle time available to the qubits, potentially increasing the noise fidelity of the circuit. Furthermore, QCLMA relies on a tree structure for generating carryout, which is more resilient to error accumulation than the single longer Carry path of Kim et al.'s quantum modulo (2\textsuperscript{n} - 1) adder that passes through all the qubits. As a result, the improved noise fidelity of the proposed QCLMA makes it a more suitable choice for NISQ era quantum computers that suffer from higher noise and erroneous results. In the fault-tolerant quantum (FTQ) era quantum computers featuring error correction, the proposed QCLMA would still be relevant as it would execute faster due to its lower depth.

To illustrate the superior noise fidelity of the proposed quantum modulo (2\textsuperscript{n} - 1) adder over Kim et al.'s version \cite{kim2021quantum}, we run our experiments on a 27-qubit quantum computer called IBM Cairo using IBM Qiskit \cite{qiskitSoftware}. Due to excessive noise in the current NISQ era quantum computers, we compare the proposed QCLMA with Kim et al.'s QRCA based modulo (2\textsuperscript{n} - 1) adder using Quantum State Fidelity Ratio (QSFR), a figure of merit which measures the ratio of the frequency of correct output to the frequency of top occurring output. We calculate the QSFR for both our proposed QCLMA and Kim et al.'s QRCA based modulo (2\textsuperscript{n} - 1) adder for all possible input combinations. Our results demonstrate that our proposed design has a 47.21\% higher QSFR than Kim et al.'s design, indicating that our design produces correct output closer to the top output than Kim et al.'s design. To summarize, our main contributions are:
\begin{itemize} [leftmargin=+.3cm]
	\item We propose a quantum carry-lookahead modulo (2\textsuperscript{n} - 1) adder (QCLMA) with an O(log n) depth compared to the O(n) depth of existing works. The proposed QCLMA offers superior noise fidelity and scalability via its improved carry path and reduced depth.
	\item We demonstrate the higher noise resilience of the proposed QCLMA over Kim et al.'s existing design by running experiments on a 27-qubit IBM Quantum Computer IBM Cairo.
	\item We define the quantum state fidelity ratio (QSFR), a figure of merit for noise resilience to measure the closeness of the correct output to the top output. Our proposed QCLMA has a 47.21\% higher QSFR in a 4-qubit configuration, establishing the effectiveness of our design.
\end{itemize}

This paper is organized with section \ref{Background} providing background on quantum gates and modulo addition. Section \ref{Related Works} discusses the existing works related to this research. Section \ref{Proposed Design} presents the proposed design. Section \ref{Noise-Resilience} compares the performance of the proposed quantum modulo (2\textsuperscript{n} - 1) adder with Kim et al.'s version\cite{kim2021quantum}. Section \ref{Discussion} discusses the results and reasons behind the superior performance of the proposed quantum modulo (2\textsuperscript{n} - 1) adder. Finally, section \ref{Conclusion} concludes this work.

\section{Background}
\label{Background}

	\subsection{Quantum Gates}
\begin{figure}[h]
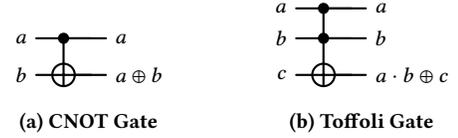

		\centering
		\small
		\begin{subfigure}[b]{0.2\textwidth}
			\centering
			\myCnot
			\caption{CNOT Gate}
			\label{fig:CNOT}
		\end{subfigure}
		\begin{subfigure}[b]{0.2\textwidth}
			\centering
			\myToffoli
			\caption{Toffoli Gate}
			\label{fig:Toffoli}
		\end{subfigure}	
		\caption{The CNOT and Toffoli gates.}
		\label{fig:qgates}
	\end{figure}

	To provide a comprehensive understanding, a detailed explanation of the quantum gates utilized in this paper is presented below.
	\begin{itemize} [leftmargin=+.3cm]
		\item CNOT Gate: CNOT gate, also known as the Feynman gate is shown in Figure~\ref{fig:qgates}(a). It maps |A,B$\rangle$ to |A,$A\oplus$B$\rangle$ helping realize XOR and XNOR operations. 
		\item Toffoli Gate: Toffoli gate, also referred to as CCNOT (double controlled NOT) gate, achieves AND and NAND logical operations depending upon value of the third input |C$\rangle$. It is the most resource intensive gate used in this work. Figure~\ref{fig:qgates}(b) shows how the  Toffoli gate maps three input |A,B,C$\rangle$ to three outputs |A,B,A$\cdot$B$\oplus$C$\rangle$.  
	\end{itemize}
	
	\subsection{Modulo (2\textsuperscript{n} - 1) addition}
	The modulo addition operation accepts two inputs, a and b, fulfilling 0 $\leq$ a,b < N, and provides (a + b) mod N as the output, where N is the modulus. The modulo operation performs division on the Sum and discards the quotient to return only the remainder. As represented by Equation \ref{equation:ModAddition}, the modulo (2\textsuperscript{n} - 1) addition is a special case of modulo addition where the modulus N = (2\textsuperscript{n} - 1). The modulo 2\textsuperscript{n} and (2\textsuperscript{n} - 1) are efficient in their binary representation, as they can represent the maximum numbers possible in n-bits. Modulo (2\textsuperscript{n} - 1) has the additional advantage of representing prime numbers depending on the value of n. Due to this unique property of modulo (2\textsuperscript{n} - 1) adders, optimizing them can help optimize various cryptography systems such as RSA that use large prime numbers \cite{ding2005rainbow, petzoldt2020efficient}. Modulus (2\textsuperscript{n} - 1) can also reduce computation when used alongside other moduli like 2\textsuperscript{n} and (2\textsuperscript{n} + 1) within a residue number system (RNS) \cite{didier2013fast}. Finally, the modulo (2\textsuperscript{n} - 1) adder can help enable other operations of the modulo and RNS algebra, such as subtraction and multiplication \cite{markov2012constant}. 
	
	\begingroup
	\setlength\abovedisplayskip{0pt}
	\begin{equation} \label{equation:ModAddition}
		\begin{aligned} 				  
			&Sum=\begin{cases}
				a + b, & \text{if}\;0\leq{(a+b)}<{2\textsuperscript{n} - 1} \\
				0, & \text{if}\;{(a+b)}={2\textsuperscript{n} - 1} \\
				(a + b)mod(2\textsuperscript{n} - 1) & \text{if}\;{(a+b)}>{2\textsuperscript{n} - 1}
			\end{cases} \\[10pt]
			&\text{where }\;0\leq{a,b,Sum}<{2\textsuperscript{n} - 1} \\
		\end{aligned}
	\end{equation}
	\endgroup
	
	However, a common problem faced by the modulo adders is the double representation of zero when zero is represented by all bits set to zero and all bits set to one. In finite field number systems used in cryptography and RNS, a double representation of zero can cause ambiguity in mathematical operations. We avoid this issue by ensuring that both the inputs and output remain within the range (0, 2\textsuperscript{n}-2).

\begin{figure*}[!h]
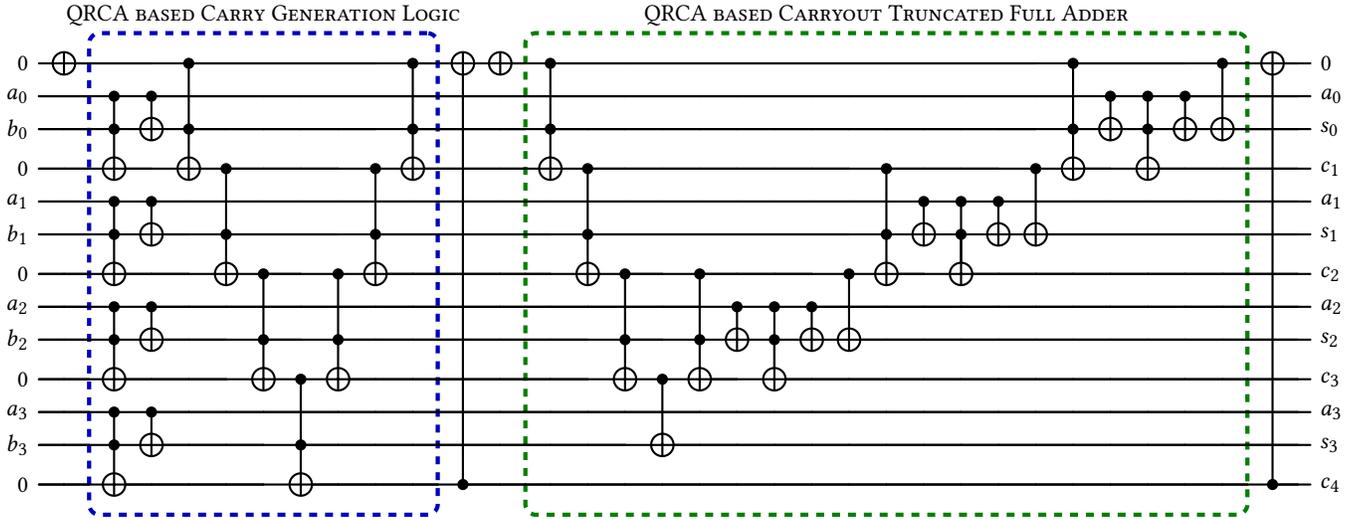

	\centering
	\KimModuloAdderBoxed
	\vspace{1mm}
	\caption{Quantum ripple-carry adder (QRCA) based quantum modulo (2\textsuperscript{n} - 1) adder by Kim et al. in 4-qubit configuration \cite{kim2021quantum}. The inputs are (a\textsubscript{0}:a\textsubscript{3}) and (b\textsubscript{0}:b\textsubscript{3}) while (S\textsubscript{0}:S\textsubscript{3}) represent Sum qubits and (c\textsubscript{0}:c\textsubscript{3}) represent the Carry qubits. The QRCA mechanism causes higher depth and longer paths in both Carry and the Sum generation, increasing the noise susceptibility. The LSB qubits are more noise prone as they spend significantly longer time idling than the MSB qubits \cite{jayashankar2022achieving}.} 
	\label{fig:KimAdder}
\end{figure*}

\section{Related Works}
\label{Related Works}
The research on quantum modulo adder began with Vedral et al.'s proposal of a general quantum modulo adder that calculates the output as (a+b) mod N, with a and b being input satisfying the condition: 0$\leq$a,b$\le$N \cite{vedral1996quantum}. Their method involves using the output of a quantum full adder and subtracting N depending on whether a+b is greater or smaller than N. Unfortunately, although their design is versatile, allowing for any integer value of modulus N, it requires five stages, making it impractical for NISQ.

Further interesting research focused on optimizing quantum modulo adders for specific applications. For instance, Roetteler et al. optimize a variant of quantum modulo adder for use in reversible elliptic curve operations in quantum cryptography, which uses a constant integer modulus \cite{roetteler2017quantum}. However, their design also suffers from high depth as it uses four stages. On the other hand, Markov and Saeedi do not design a modulo adder but instead propose a reversible circuit for conditional modular addition with a constant as they focus on optimizing modular multiplication for Shor's quantum number-factoring algorithm \cite{markov2012constant}.

Additional research on quantum modulo adders accompanied the development of quantum full adders. For quantum modulo 2\textsuperscript{n} adders, Cuccaro et al. and Takahashi et al. extend their designs for the quantum full adders to quantum modulo 2\textsuperscript{n} adders using simple truncation of the carryout \cite{cuccaro2004new, takahashi2005linear}. This technique works as an n-bit output is expected from n-bit input addition, with the truncated carryout being the most significant bit (MSB) representing the 2\textsuperscript{n} position.

Draper et al. propose a new architecture for quantum carry-lookahead adders with O(log n) depth and extend their design to both modulo 2\textsuperscript{n} and modulo (2\textsuperscript{n} - 1) adders \cite{draper2004logarithmic}. However, their research lacks sufficient information on the conversion process and its limitations. Additionally, their design uses one's complement arithmetic, which represents zero with both all zeros and all ones, creating the problem of double representation of zero that is not addressed in their research.

Kim et al. design a quantum modulo adder for the Galois Field GF(2\textsuperscript{n} - 1) useful in ECC and multivariate quadratic-based post-quantum cryptography (MQPQC). Their design outputs (a+b) mod (2\textsuperscript{n} - 1) with the inputs satisfying condition: 0$\leq$ a,b <(2\textsuperscript{n} - 1) \cite{kim2021quantum, ding2005rainbow, petzoldt2020efficient}.

\begin{subnumcases}{Sum=}
	0, & if (a+b)=2\textsuperscript{n} - 1 \label{equation:KimAdd1}
	\\
	(a + b + c\textsubscript{n})mod(2\textsuperscript{n} - 1) & else \label{equation:KimAdd2}
\end{subnumcases}
\vspace{3mm}
\indent\indent\hspace{0.3em}\text{where }0$\leq a,b,Sum $ < {2\textsuperscript{n} - 1}

Their approach works in two stages. First, they implement Equation \ref{equation:KimAdd2} using Carry generation logic to calculate carryout c\textsubscript{n} and feed it as input Carry to Carry truncated full adder performing addition of a and b. As shown in Figure \ref{fig:KimAdder}, their optimized circuit utilizes two ripple carry addition units, first for generating Carry and second for Carry truncated addition. However, if the Sum is equal to the modulus, the set zero logic acts to return zero. Figure \ref{fig:KimAdder} shows the set zero logic enabled using two CNOT and two NOT gates outside the boxes denoting the addition units. The set zero logic also solves the problem of double representation of zero. Our upcoming section will detail our proposed design, which leverages our comprehension of the approach adopted by the aforementioned research.

\section{Proposed Design}
\label{Proposed Design}

\begin{figure*}[!h]
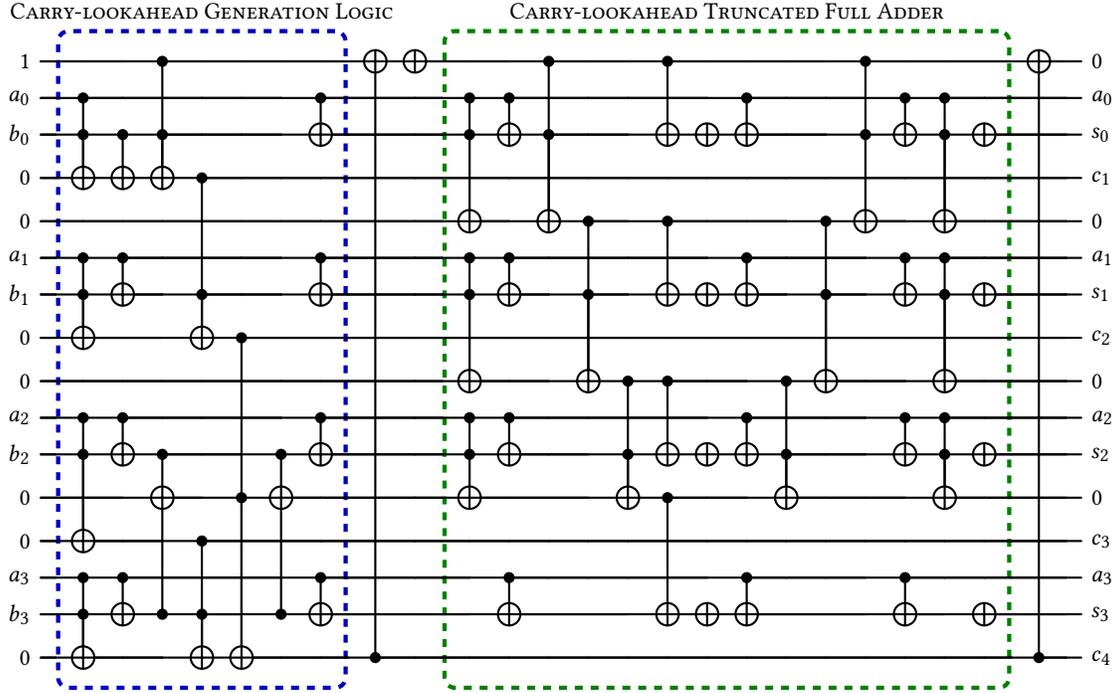

	\centering
	\ProposedModuloAdderBoxed
	\caption{Proposed quantum carry-lookahead modulo adder (QCLMA) in 4-qubit configuration. First, the Carry generation logic takes the inputs (a\textsubscript{0}:a\textsubscript{3}) and (b\textsubscript{0}:b\textsubscript{3}) to generate Carry qubits (c\textsubscript{0}:c\textsubscript{4}). The out-of-place carry-lookahead generation logic makes sure that the inputs are passed intact for next stage. The in-place carry-lookahead truncated full adder generates the Sum (S\textsubscript{0}:S\textsubscript{3}) in place of input b. Both the Carry and Sum are generated using a tree-based structure which has a shorter O(log n) depth. Lesser variation in idling time among the qubits makes the proposed QCLMA noise resilient.} 
	\label{fig:ProposedAdder}
\end{figure*}

\renewcommand{\arraystretch}{1.2}
\begin{table*}[h]
	\caption{\textsc{Quantum Resources for Proposed QCLMA and Kim et al.'s QRCA based Modulo (2\textsuperscript{n} - 1) Adders}\\ ($\mathbf{\log_2}$ denoted as log and w(n) is number of 1's in binary representation of n.)}
	\label{table:resources}
	\centering
\begin{tabular}{|l|l|cc|ccc|c|}
	\hline
	\multirow{2}{*}{\textbf{Adder}} &
	\multirow{2}{*}{\textbf{Sub-Modules}} &
	\multicolumn{2}{c|}{\textbf{Depth}} &
	\multicolumn{3}{c|}{\textbf{Gate Count}} &
	\multirow{2}{*}{\textbf{Qubits}} \\ \cline{3-7}
	&
	&
	\multicolumn{1}{c|}{\textbf{CNOT}} &
	\textbf{Toffoli} &
	\multicolumn{1}{c|}{\textbf{CNOT}} &
	\multicolumn{1}{c|}{\textbf{Toffoli}} &
	\textbf{NOT} &
	\\ \hline
	\multirow{4}{*}{\textbf{Kim et al.}} &
	Carry Generator &
	\multicolumn{1}{c|}{1} &
	2n &
	\multicolumn{1}{c|}{n} &
	\multicolumn{1}{c|}{3n - 1} &
	0 &
	\multirow{3}{*}{} \\ \cline{2-7}
	&
	Truncated Adder &
	\multicolumn{1}{c|}{3n - 2} &
	3n - 3 &
	\multicolumn{1}{c|}{3n - 2} &
	\multicolumn{1}{c|}{3n - 3} &
	0 &
	\\ \cline{2-7}
	&
	Set Zero Logic &
	\multicolumn{1}{c|}{2} &
	0 &
	\multicolumn{1}{c|}{2} &
	\multicolumn{1}{c|}{0} &
	2 &
	\\ \cline{2-8} 
	&
	Total &
	\multicolumn{1}{c|}{3n + 1} &
	5n - 3 &
	\multicolumn{1}{c|}{4n} &
	\multicolumn{1}{c|}{3n -4} &
	2 &
	3n+1 \\ \hline
	\multirow{4}{*}{\textbf{\begin{tabular}[c]{@{}l@{}}Proposed \\ QCLMA\end{tabular}}} &
	Carry Generator &
	\multicolumn{1}{c|}{2} &
	$\lfloor log(n) \rfloor$+3 &
	\multicolumn{1}{c|}{2n} &
	\multicolumn{1}{c|}{4n-3w(n)-2$\lfloor log(n) \rfloor$+1} &
	0 &
	\multirow{3}{*}{} \\ \cline{2-7}
	&
	Truncated Adder &
	\multicolumn{1}{c|}{4} &
	$\lfloor log(n) \rfloor$+2$\lfloor log(n-1) \rfloor$+2 &
	\multicolumn{1}{c|}{4n} &
	\multicolumn{1}{c|}{4n-4} &
	2n &
	\\ \cline{2-7}
	&
	Set Zero Logic &
	\multicolumn{1}{c|}{2} &
	0 &
	\multicolumn{1}{c|}{2} &
	\multicolumn{1}{c|}{0} &
	1 &
	\\ \cline{2-8} 
	&
	Total &
	\multicolumn{1}{c|}{8} &
	3$\lfloor log(n) \rfloor$+2$\lfloor log(n-1) \rfloor$+5 &
	\multicolumn{1}{c|}{6n+2} &
	\multicolumn{1}{c|}{8n-3w(n)-2$\lfloor log(n) \rfloor$-3} &
	2n+1 &
	3n+2+2$\lfloor log(n) \rfloor$ \\ \hline
\end{tabular}
\end{table*}

In the previous section, we discussed how Kim et al. designed the quantum modulo (2\textsuperscript{n} - 1) adder. However, as both stages in Kim et al.'s quantum modulo (2\textsuperscript{n} - 1) adder uses quantum ripple carry addition (QRCA) logic, it suffers from issues persistent in Vedral et al. For example, the depth grows linearly with input qubit size, causing the Carry propagation chain to become longer, thereby reducing the noise fidelity. We address these issues in the proposed quantum carry-lookahead modulo (2\textsuperscript{n} - 1) adder (QCLMA) using by introducing logarithmic depth of O(log n) complexity. We also utilize tree based Carry propagation which is more resilient to noise than the chain based Carry propagation of QRCA.

To construct the proposed quantum carry-lookahead modulo (2\textsuperscript{n} - 1) adder (QCLMA), we selectively choose between two types of logic: out-of-place and in-place. The out-of-place logic passes both the inputs unchanged and generates Sum on ancillae. As depicted in Figure \ref{fig:ProposedAdder}, we use it as the first stage of our quantum modulo (2\textsuperscript{n} - 1) adder as it can pass the inputs to be used by the second stage. We optimize the out-of-place logic to generate only Carry while passing the inputs to second stage. Furthermore, we initialize the first ancilla with |1$\rangle$ to eliminate the redundant NOT gate.

The first stage generates carryout c\textsubscript{n} for the second stage, which uses it as input Carry for carryout truncated full adder, like in Equation \ref{equation:KimAdd2}. Here, we use an in-place quantum carry-lookahead adder which generates Sum in place of input B saving ancillae, as evident from Figure \ref{fig:ProposedAdder}. The set zero logic mechanism works in the same manner as before. Table \ref{table:resources} shows the quantum resource usage and depth for the two adders. We see an increase in qubits used in the proposed QCLMA as it uses extra qubits as ancillae. The depth, on the other hand, is O(log n) compared O(n) of Kim et al.'s quantum modulo (2\textsuperscript{n} - 1) adder. 

For a 4-qubit configuration, the proposed QCLMA has 38\% lower CNOT depth compared to Kim et al.'s modulo (2\textsuperscript{n} - 1) adder. Here CNOT depth is defined as a measure of CNOT gate layers that can be executed in parallel in the quantum circuit. Figures \ref{fig:KimAdder} and \ref{fig:ProposedAdder} illustrates that the CNOT depth in the proposed QCLMA and Kim et al.'s version are eight and thirteen, respectively. Similarly, the proposed QCLMA has 23\% lower Toffoli depth compared to Kim et al.'s modulo (2\textsuperscript{n} - 1) adder, as the Toffoli depth in proposed QCLMA and Kim et al.'s version are thirteen and seventeen, respectively. We define Toffoli depth as a measure of Toffoli gate layers that can be executed in parallel in the quantum circuit. In the section \ref{Noise-Resilience}, we present a detailed analysis of the implications of depth reduction on noise resilience.

\section{Noise-Resilience of proposed design}
\label{Noise-Resilience}
To compare the noise resilience of the proposed quantum modulo (2\textsuperscript{n} - 1) adder with Kim et al.'s quantum modulo (2\textsuperscript{n} - 1) adder, we run experiments on a 27-qubit IBM quantum computer (IBM Cairo) using the following methodology. We first create 4-qubit versions of both circuits as shown in Figure \ref{fig:KimAdder} and Figure \ref{fig:ProposedAdder}. For four bits, the modulus is (2\textsuperscript{4} - 1) = 15, and the range of input and output values is $0 \le a, b, Sum < 15$.

For the proposed quantum adder, we fix the value of input A as constant but iterate over the value of B from (0, 14) to create fifteen circuits appended to a single job before being sent to IBM Cairo. We batch fifteen such jobs for each value of A in (0, 14) to run the proposed quantum adder for all 255 input combinations. We use the same methodology also to generate and batch the jobs for Kim et al.'s quantum adder. All jobs are dispatched with 1024 shots. The output from IBM Cairo is received in the form of shot frequency in the sixteen different possible outputs, from 0000 to 1111.

Since the output of NISQ era machines such as IBM Cairo is noisy, the correct output may not stand among the top three outputs and strongly depends on the input combinations. For example, for the input combination causing the least delay: A = B = 0, the correct output is the output with the highest shot count for both quantum modulo (2\textsuperscript{n} - 1) adders. Whereas for the input combination causing the maximum delay: A = B = 14, the correct output is at a fourteenth place for Kim et al.'s modulo adder and sixth place for the proposed modulo adder. Since the rank of correct output among all the possible outputs does not measure it's degree of divergence from the top output, we need another criterion to facilitate a meaningful comparison of the adders.

Hence we design a figure of merit called Quantum State Fidelity Ratio (QSFR) to measure the closeness of the correct output to the top output. As shown in Equation \ref{equation:QSFR}, QSFR is the ratio of the frequency of the correct output to the frequency of the top output. QSFR helps us to quantify the fidelity and degree of improvement needed to advance the correct answer to the top position.
\vspace{1mm}
\begin{equation} \label{equation:QSFR}
	\text{QSFR} = \frac{\text{Frequency of Correct Output}}{\text{Frequency of Top Output}}
	\vspace{3mm}
\end{equation}

\begin{figure}[h]\captionsetup[subfigure]{font=footnotesize}
	\centering
	%\vspace{-1mm}
	\includegraphics[width=\columnwidth]{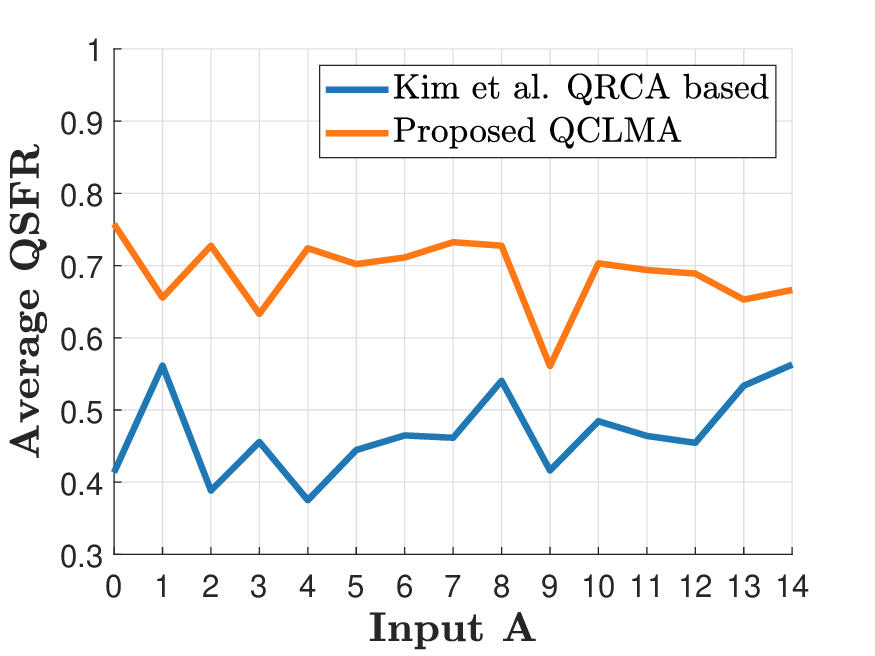}
	%\vspace{-1mm}
	\setlength{\abovecaptionskip}{-7pt}
	\setlength{\belowcaptionskip}{-23pt}
	\caption{Comparison of Average QSFR of Proposed QCLMA and Kim et al.'s QRCA based modulo (2\textsuperscript{n} - 1) adder, obtained keeping constant A and varying B. Proposed adder has 47.21\% higher average QSFR establishing superior noise fidelity.}
	\label{fig:FOM}
	\vspace{3mm}
\end{figure}

We calculate the average QSFR for every value of input A added with B in range (0, 14) and plot them in Figure \ref{fig:FOM}. Each tick on the x-axis represents the QSFR averaged for a job with constant input A and input B iterating in range (0, 14). The mean of all the average QSFR helps quantify the improvement needed by the design considering all 255 input combinations, which computes to 0.6891 for the proposed QCLMA compared to 0.4681 for the Kim et al.'s QRCA based modulo (2\textsuperscript{n} - 1) adder. The substantial 47.21\% rise in average QSFR confirms the superior noise fidelity and enhanced potential for improvement in the proposed QCLMA. In the forthcoming section, we analyze various factors contributing to this improvement. 

\begin{figure*}[htbp]
	\centering
	\begin{subfigure}[b]{0.45\textwidth}
		\centering
		%\hspace{-2mm}
		\includegraphics[width=1.15\columnwidth]{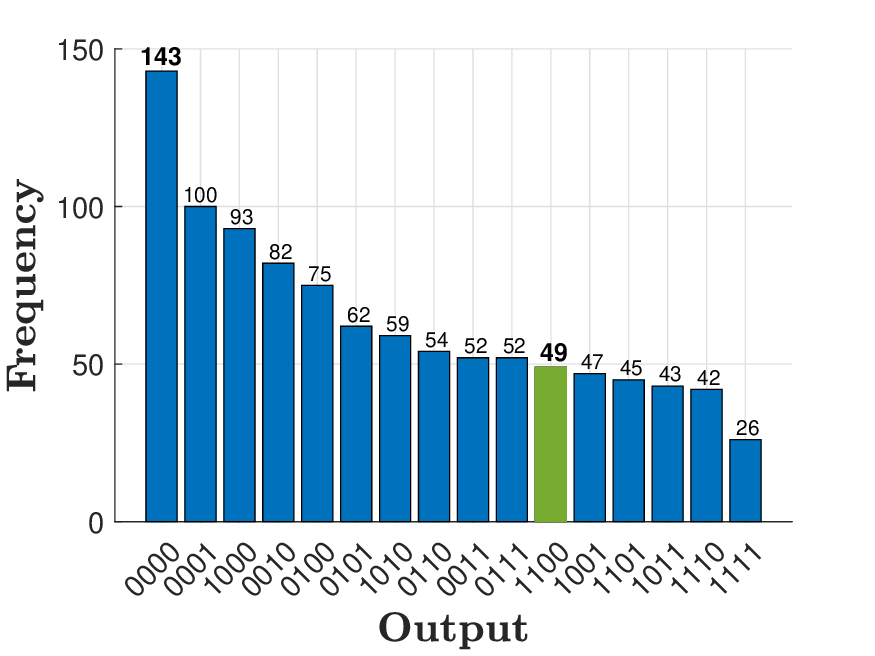}
		\caption{Kim et al.'s QRCA based Modulo (2\textsuperscript{n} - 1) adder}
		\label{fig:kim102}
	\end{subfigure}
	%\hspace{4mm}
	\hfill
	\begin{subfigure}[b]{0.45\textwidth}
		\centering
		\includegraphics[width=1.15\columnwidth]{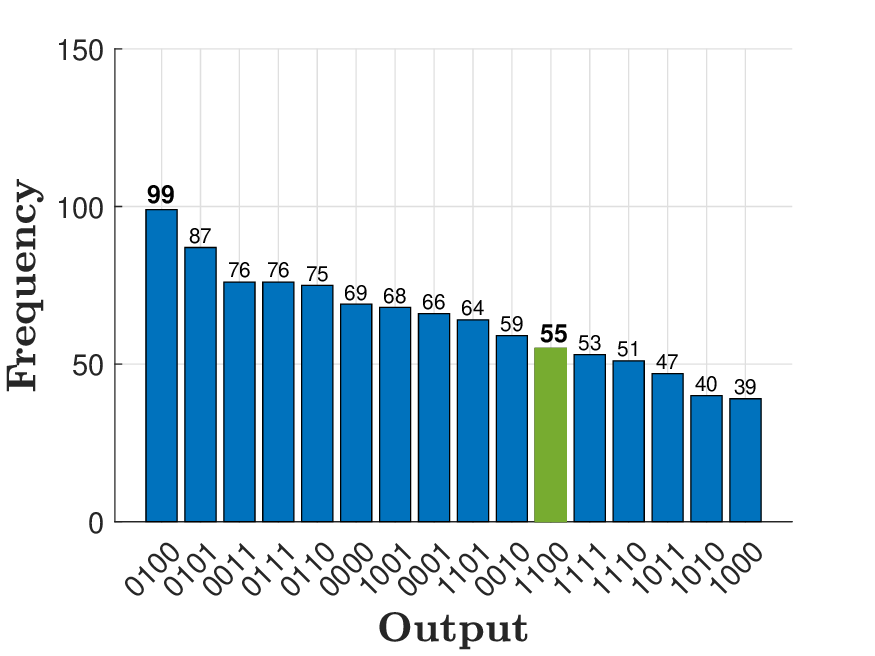}
		\caption{Proposed quantum carry-lookahead modulo adder (QCLMA)}
		\label{fig:cla102}
		%\hspace{2mm}
	\end{subfigure}
	\caption{The above graphs illustrate the output of the modulo adders generated from IBM Cairo, arranged in descending order of frequency. Identical input A = 10 and B = 2 is supplied, with both adders yielding S = 12 or 1100 as the output at the same 11\textsuperscript{th} position among 16 possible outputs. The output profile of our proposed QLCMA adder exhibits a slower descent rate than the version proposed by Kim et al., reducing the relative gap with the top output and improving its chances of improvement.}
	\label{fig:both102}
\end{figure*}

%\vspace{-5mm}
\section{Discussion}
\label{Discussion}
To gain further insight into the superior performance of the proposed quantum carry-lookahead modulo adder (QCLMA), we examine an input combination of A = 10 and B = 2. In this scenario, both modulo (2\textsuperscript{n} - 1) adders yield the output S = 12 at the eleventh position among sixteen possible outputs. Figure \ref{fig:both102} highlights the variation in QSFR between the two cases, despite the frequency of the correct output being nearly identical. For instance, Figure \ref{fig:both102}(b) shows the proposed QCLMA has the correct output with a frequency of 55 and QSFR of 0.55, compared to the frequency of 49 and QSFR of 0.343 for Kim et al.'s QRCA based modulo (2\textsuperscript{n} - 1) adder shown in Figure \ref{fig:both102}(a).

Figure \ref{fig:both102}(a) displays the output frequency profile of Kim et al.'s QRCA based modulo (2\textsuperscript{n} - 1) adder, revealing a nearly exponential decline from the top output. It means that the correct output unless it is the top occurring one, will have an exponentially increasing gap with the top output. The source of this exponential drop is the QRCA architecture that introduces a considerable variation in the depth and qubit idling time. In comparison, Figure \ref{fig:both102}(b) shows that the proposed QCLMA follows a linear decrease in frequency under the same experimental conditions, demonstrating a higher resilience to disruptions caused by noise. QCLMA's tree structure essentially creates a self-normalization effect, reducing inter-output variation and introducing noise robustness.

\section{Conclusion}
\label{Conclusion}
In this work, we have proposed quantum carry-lookahead modulo (2\textsuperscript{n} - 1) adder (QCLMA) that primarily introduces O(log n) depth compared to O(n) depth of existing work. We demonstrate the impact of O(log n) depth in improving noise resilience by running experiments on a 27-qubit quantum computer called IBM Cairo. As a result, the proposed QCLMA improves by 47.21\% in terms of noise fidelity using quantum state fidelity ratio (QSFR), a figure of merit that measures the closeness of the correct output to the top output. We also trace the origin of this performance by comparing the output frequency profile of the proposed quantum modulo (2\textsuperscript{n} - 1) adder with existing work obtained under identical input and output scenarios.

Our results illustrate that the shorter depth and distributed Carry path increase the noise fidelity and make the QCLMA suitable for NISQ era quantum computers. The proposed QCLMA can be instrumental in creating quantum modulo subtraction and multiplication circuits, and for applications such as quantum cryptography and quantum image processing. We conclude that QCLMA has the potential to benefit quantum modulo arithmetic circuits by increasing their noise resilience. 

%%
%% The acknowledgments section is defined using the "acks" environment
%% (and NOT an unnumbered section). This ensures the proper
%% identification of the section in the article metadata, and the
%% consistent spelling of the heading.
%\begin{acks}
%\end{acks}

%%
%% The next two lines define the bibliography style to be used, and
%% the bibliography file.
\bibliographystyle{ACM-Reference-Format}
\bibliography{sample-base}

%%
%% If your work has an appendix, this is the place to put it.

\end{document}